\documentclass{sf2a-conf2018}
\usepackage{graphicx}
\usepackage{hyperref}
\usepackage[]{natbib}  
\usepackage{epstopdf}
\usepackage{pgf, tikz}
\usepackage{amsmath,amssymb}
\usepackage{soul}

\def\BibTeX{{\rm B\kern-.05em{\sc i\kern-.025em b}\kern-.08em
    T\kern-.1667em\lower.7ex\hbox{E}\kern-.125emX}}
\bibpunct{(}{)}{;}{a}{}{,}  

\begin{document}

\TitreGlobal{SF2A 2018}


\title{The effects of rotation on wave-induced transport in stars: from weakly to strongly stratified radiative zones}

\runningtitle{Rotation effects on wave-induced angular momentum transport}

\author{Q. Andr\'e}\address{AIM, CEA, CNRS, Universit\'e Paris-Saclay, Universit\'e Paris Diderot, Sorbonne Paris Cit\'e, F-91191 Gif-sur-Yvette, France}
\author{S. Mathis$^{1,}$}\address{LESIA, Observatoire de Paris, CNRS UMR 8109, UPMC, Universit\'e Paris-Diderot, 5 place Jules Janssen, 92195 Meudon, France}
\author{L. Amard}\address{University of Exeter, Department of Physics \& Astronomy, Stoker Road, Devon, Exeter, EX4 4QL, UK}

\setcounter{page}{237}


\maketitle


\begin{abstract}
Internal waves propagating in stellar radiative zones can lead to efficient angular momentum transport{, that should occur throughout the whole lifetime of stars.} 
They thus play a key role in  {shaping} the internal rotation profile of these regions, that can be probed by asteroseismology. We present a new analytical study of their propagation and dissipation  {near the equatorial plane.} {We include} the effects of rotation and differential rotation
without making any assumption on their relative strength relative to that of the background stable stratification. This analytical framework allows in principle to scan the efficiency of the wave-induced transport of angular momentum. {The computations goes} from the pre-main sequence, during which the restoring forces associated with rotation and stratification can be of the same order, to the later stages of evolution, for which stratification tends to dominate over rotation. A {first} application to the case of a sun-like star is finally discussed.
\end{abstract}

\begin{keywords}
Hydrodynamics, Waves, Methods: analytical, Stars: interiors, Stars: evolution, Stars: rotation, Stars: low-mass
\end{keywords}


\section{Context}
Thanks to asteroseismic measurements, internal rotation profiles of stars have become an  {observed stellar property \citep[for a recent review see e.g.][]{AertsEtal2018}.} 
For instance, it has been shown that the rotation rate in the radiative zone (RZ) of the Sun is uniform
 {down to} $0.2R_{\odot}$ \citep[e.g.][]{GarciaEtal2007,FossatEtal2017}, that cores of red giants are the seat of a strong extraction of angular momentum along their evolution \citep[e.g.][]{MosserEtal2012,BeckEtal2012,DeheuvelsEtal2014,DeheuvelsEtal2017,GehanEtal2018}, and that radiative zones of intermediate-mass and massive stars undergo weak differential rotation \citep[e.g.][]{KurtzEtal2014,SaioEtal2015,DeheuvelsEtal2015,MurphyEtal2016,VanReethEtal2016}. Supported by theoretical studies, those observations reveal that efficient mechanisms capable of  {redistributing} 
angular momentum are at work all along the evolution of stars. Among them, several studies have proposed that 
internal waves \citep[excited at the boundary with {turbulent} convective zones, see e.g.][]{AlvanEtal2014}, can be an efficient process that is able to extract angular momentum {from the} 
radiative zone in which they propagate, {to release it in the convective envelope} \citep[e.g.][]{Schatzman1993,ZahnEtal1997,TalonCharbonnel2005,FullerEtal2014,PinconEtal2017}.

State-of-the-art analytical prescriptions implemented in one-dimensional (1D) stellar evolution codes have {generally} been derived in the framework of pure gravity waves, thus neglecting the action of the Coriolis acceleration upon the propagation, dissipation and associated angular momentum transport {by} 
those waves. In reality, stars are rotating and the Coriolis acceleration provides an additional restoring force that adds to the buoyancy force. When the Coriolis frequency, $2\Omega$, is of the same order as the buoyancy frequency, $N$, we expect  {rotation to play an important role.}

Several works have considered the effect of rotation in the framework of the traditional approximation \citep[in which the latitudinal component of the rotation vector is neglected, e.g.][]{PantillonEtal2007,MathisEtal2008,Mathis2009}.  {But this is valid only} when $N/2\Omega \gg 1$. In the case of low-mass stars, this assumption is appropriate during the main sequence (MS), but not necessarily during the pre-main sequence (PMS), because then these stars can rotate fast enough so that $N/2\Omega \sim 1$, before wind braking kicks in. Besides, \cite{CharbonnelEtal2013} showed that the extraction of angular momentum by internal gravity waves plays an important role during the PMS, than can significantly affect the internal rotation profile at arrival on the zero age main sequence (ZAMS). 

Here, we propose a new analytical study that aims at taking into account rotation and radial differential rotation, for any ratio $N/2\Omega$. As a first step, we compute how they affect the  {wave} penetration length of internal waves, that is the characteristic length over which an internal wave can exchange angular momentum with the background flow before being completely damped by thermal diffusion. We highlight the main steps of our analysis in Section \ref{sec:sec1}, and show an application to a solar-like model in Section \ref{sec:sec2}. Finally, we give our  {preliminary} conclusions and prospects in Section \ref{sec:sec3}.

\section{Statement of the problem}
\label{sec:sec1}

\subsection{Equations in an equatorial model}
To carry out our analysis, we have chosen to write our equations near the equator, inspired by the approach first introduced by \cite{Ando1985}. We note that such a  {(quasi)} two-dimensional model has proven itself useful to reproduce asteroseismic constraints with numerical simulations \citep{Rogers2015}. In addition, it allows to  {filter, as a first step,}  more complex three-dimensional behaviours such as latitudinal trapping  {in the presence of (differential) rotation} \citep[e.g.][]{Prat2018}.

We write our equations for the perturbed motions in the framework of the Cowling \citep[variations of the gravitational potential by the waves are neglected, see][]{Cowling1941}, and the anelastic (sound waves are filtered out) approximations, and we include thermal diffusion. Viscosity is neglected over thermal diffusion, as {their ratio}
is expected to be very small in stellar radiative zones \citep[we refer the reader to Fig. 1 in][]{BrunZahn2006}. 
{The stellar angular velocity is $\Omega(r)$.}
We thus write the following set of equations on the velocity, density  {$\rho$} and pressure  {$P$} perturbations:
\begin{gather}
i\omega v_r - 2\Omega v_{\phi} = -\frac{\text{d}W'}{\text{d}r} + \frac{\rho'}{\rho^2} \frac{\text{d}P}{\text{d}r},\\
i\omega v_{\theta} = -\frac{ikW'}{r},\\
i\omega v_{\phi} + \zeta v_r = -\frac{imW'}{r},\\
\frac{\text{dln}\rho}{\text{d}r}v_r + \frac{1}{r^2} \frac{\text{d}(r^2v_r)}{\text{d}r} + \frac{ikv_{\theta}}{r} + \frac{imv_{\phi}}{r} = 0,\\
i\omega\frac{\rho'}{\rho} = \frac{N^2}{g_e}v_r + \kappa \nabla^2\left(\frac{\rho'}{\rho}\right),
\end{gather}
which correspond  {respectively} to the three components of the linearised equation of motion, continuity equation, and {heat transport} equation. In addition, $v_r$, $v_{\theta}$, $v_{\phi}$ are the radial, latitudinal and azimuthal components of the perturbed velocity, $W'=P'/\rho$, $\zeta = 2\Omega + r\text{d}\Omega/\text{d}r$, $N^2$ is the buoyancy frequency, {$\kappa$ the thermal diffusivity,} and $g_{\text{e}}$ is the  {local} effective gravity (including the centrifugal acceleration). The perturbed parts of each unknown are assumed to vary as $\exp\left\{ i (\sigma t + k\theta + m\phi)\right\}$.  {We have introduced the Doppler-shifted} frequency $\omega = \sigma + m\Omega$, where $\sigma$ is the frequency of excitation and $m$ the azimuthal wave number.

\subsection{Local dispersion relation}
Following \cite{Ando1985}, we consider a short-wavelength wave in the radial direction and apply the WKBJ approximation, so that perturbed parts will also  {vary} as $\exp\left\{ i \int k_r \text{d}r\right\}$. Keeping leading order terms, we get the following dispersion relation, that includes thermal diffusion:
\begin{equation}
    \left(\frac{N^2 + 2\Omega\zeta\alpha}{\omega^2} -1\right)k_h^2 - k_r^2 = \frac{i\kappa k^2}{\omega} \left\{ \left( \frac{2\Omega\zeta\alpha}{\omega^2} -1 \right)k_h^2 - k_r^2 \right\},
      \label{eq:disprel}
\end{equation}
where $k_h^2 = l^2/r^2$, $l^2 = k^2 + m^2$ and 
\begin{equation}
    \alpha = 1 - \left(\frac{m}{l}\right)^2.
      \label{eq:alpha}
\end{equation}

We then carry out a quasi-adiabatic approach similar to \cite{Press1981} and \cite{ZahnEtal1997}, relevant for stellar radiative zones in which thermal diffusion can be treated as a small effect. In the quasi-adiabatic limit, radiative damping provides an attenuation factor $\exp(-\tau/2)$ of the wave's amplitude. The expression of $\tau$ follows from injecting the expression $k_r = k_{r,\text{ad}} + ik_{r,\text{diss}}$ into the dispersion relation above, assuming that $k_{r,\text{diss}} \ll k_{r,\text{ad}}$, and writing $\tau = \int k_{r,\text{diss}} \, \text{d}r$. 

From the equation above (setting $\kappa=0$), we first get that the adiabatic radial wave number writes
\begin{equation}
    k_{r,\text{ad}}^2 = k_h^2 \left(\frac{N^2 + 2\Omega\zeta \alpha}{\omega^2} - 1\right).
\end{equation}
 {We recognize the terms associated to buoyancy, rotation, and wave acceleration, respectively.}
Then, carrying out the quasi-adiabatic analysis, we get that the damping factor is the integral
\begin{equation}
    \tau = \int_r^{r_{\text{c}}} \left(\frac{\kappa k_h^2}{N}\right)\left(\frac{N}{\omega}\right)^2\left(\frac{N^2 + 2\Omega\zeta\alpha}{\omega^2}\right)\left(\frac{N^2}{N^2+2\Omega\zeta\alpha-\omega^2}\right)^{1/2}k_h\, \text{d}r,
\end{equation}
 {where $r_{\text{c}}$ is the radius of the radiative/convective boundary, here in the case of a convective envelope.} The expression above can be compared to the one for pure gravity waves \citep[e.g.][]{ZahnEtal1997}, given by
\begin{equation}
    \tau_0 = \int_r^{r_{\text{c}}} \left(\frac{\kappa k_h^2}{N}\right)\left(\frac{N}{\omega}\right)^4\left(\frac{N^2}{N^2-\omega^2}\right)^{1/2}k_h\, \text{d}r.
\end{equation}

\subsection{Derivation of the penetration length}
The expression of the damping factor can be written involving a characteristic length, $L$, such that
\begin{equation}
    \tau = \int_r^{r_{\text{c}}} \frac{\text{d}r}{L},
\end{equation}
as in \cite{FullerEtal2014}. This so-called {penetration length} provides a first proxy that we found worth focusing on  {to examine the effects of rotation and differential rotation upon the wave-induced transport.}
This represents the characteristic length over which a wave will be able to exchange angular momentum with the background flow, before being damped out by thermal diffusion. It is thus intrinsically linked to the efficiency of the coupling between the excitation region (convective zones) and radiative layers in which the waves propagate.

Let us introduce the following set of dimensionless  {Froude and Richardson} numbers:
\begin{equation}
\text{Fr} = \frac{\omega}{N}, ~~~~~ \text{S} = \frac{2\Omega}{N}, ~~~~~ \text{and} ~~~~~ \text{Ri} = \left(\frac{N}{r\text{d}\Omega/\text{d}r}\right)^2.
\end{equation}

The expression of the penetration length, normalised by that of pure gravity waves $L_0 = L(\text{S}=0,\text{Ri} = \infty)$, is thus given by
\begin{equation}
    \frac{L}{L_0} = \frac{\left(\displaystyle 1 + \alpha \frac{\mathrm{S}(\mathrm{S}+\mathrm{Ri}^{-1/2})}{1-\mathrm{Fr}^2}\right)^{1/2}}
        {\displaystyle 1 + \alpha \, \mathrm{S}(\mathrm{S}+\mathrm{Ri}^{-1/2})}.
      \label{eq:L}
\end{equation}
{From the expression above, one can see that the parameter $\alpha$, which expression is given by Eq. (\ref{eq:alpha}), somewhat weights the terms linked to rotation.  Waves with $l=m$ behave like pure gravity waves in the framework of our analysis, because then $\alpha = 0$ and thus $L=L_0$. Therefore, because waves with $m=0$ do not lead to a net transport of angular momentum, we rather expect waves with $m=1$ and $l>1$ to be the more impacted by rotation.}  

\section{Application to a sun-like star}
\label{sec:sec2}
We now explore how this ratio $L/L_0$ varies along the evolution of a $1M_{\odot}$ sun-like star.
\subsection{Description of the evolution models}
\begin{figure}
     \centering
     \includegraphics[width=0.495\linewidth]{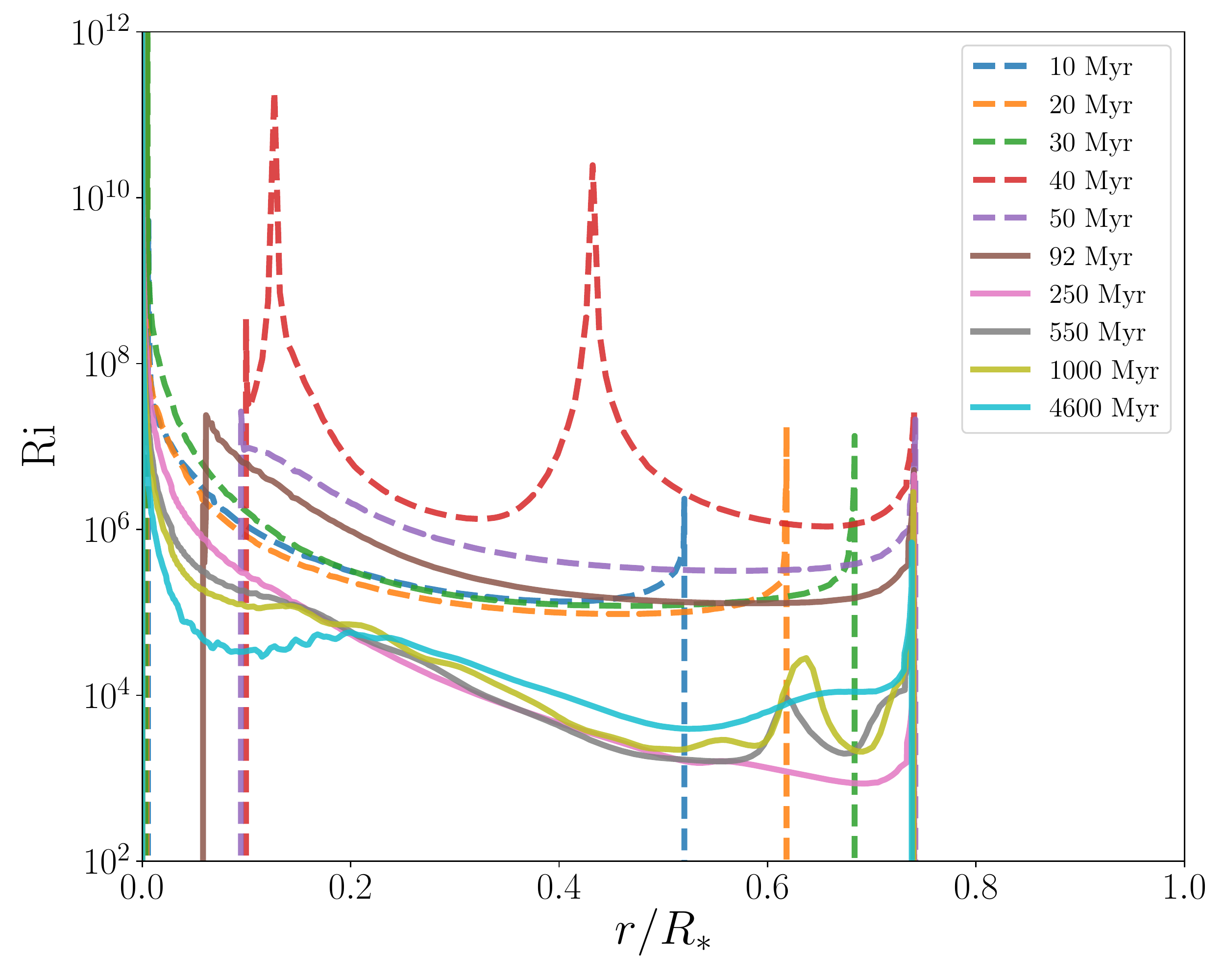}
     \includegraphics[width=0.495\linewidth]{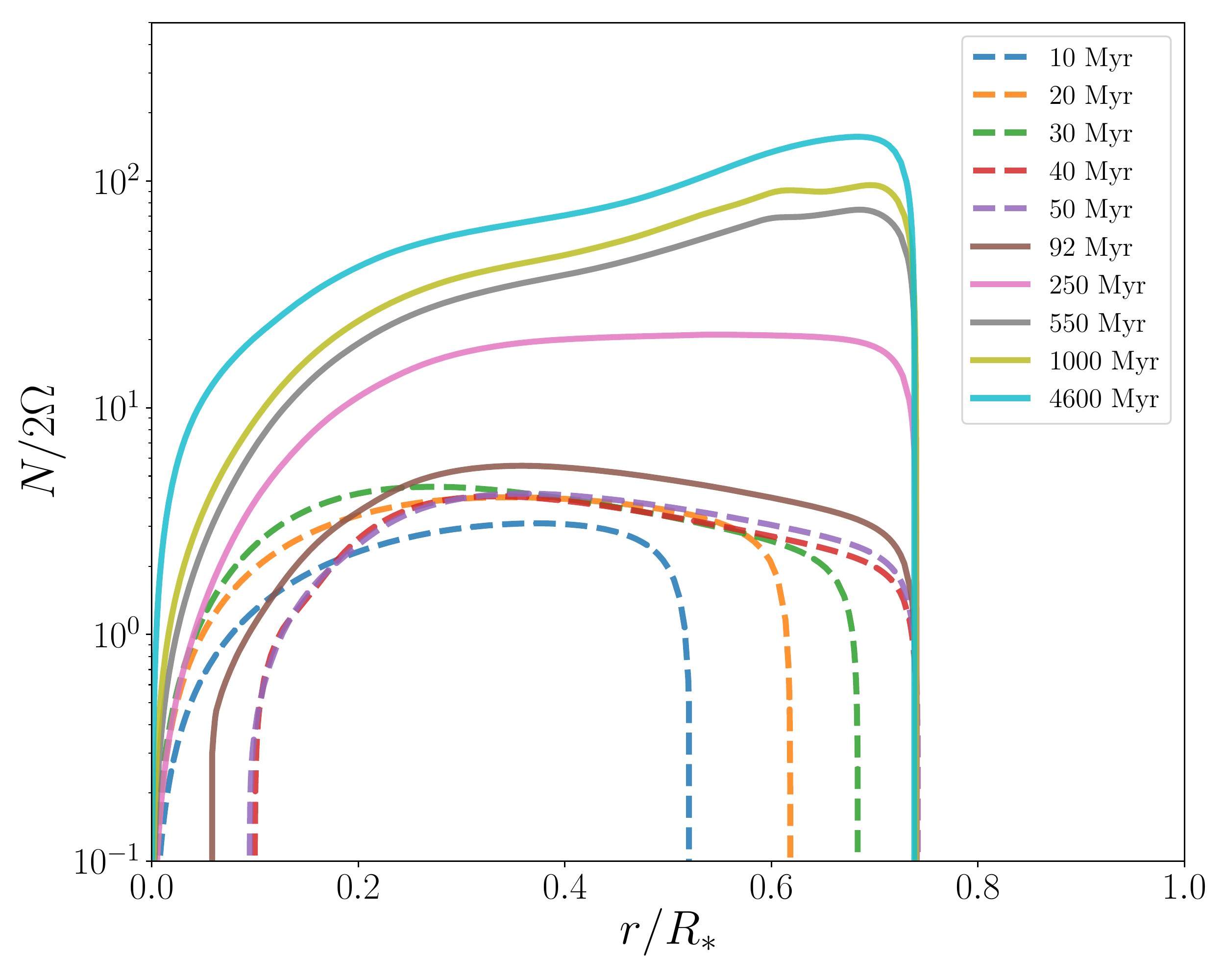}\\[-10pt]
     \caption{Richardson number Ri (\textit{left panel}) and ratio N/$2\Omega$ (\textit{right panel}) as a function of the radius normalised by the stellar radius $R_*$ at specific ages, scanning from 10 Myr to 4.6 Gyr, obtained with the stellar evolution code STAREVOL. The ZAMS occurs at 56 Myr in the simulation. On both panels, dashed lines correspond to profiles on the PMS, and solid lines correspond to profiles on the MS}
       \label{fig:models}
\end{figure}
In order to calculate realistic values of $L$, we have computed a one-dimensional evolution model of a sun-like star using the stellar evolution code STAREVOL \citep[e.g.][Amard et al. in prep]{SiessEtal2000,PalaciosEtal2003,TalonCharbonnel2005,DecressinEtal2009,LagardeEtal2012,CharbonnelEtal2013,AmardEtal2016}. { It computes angular momentum evolution (internal transport and surface extraction) in a self-consistent way during stellar evolution. However, it} only takes into account angular momentum transport due to meridional circulation and shear instabilities.  {This allows to isolate the needed effect of internal waves (and potential other missing transport processes), since these mechanisms are not sufficient on their own to reproduce the observations \citep[][]{EggenbergerEtal2012,MarquesEtal2013,CeillierEtal2013,MathisEtal2018}.}
 {To maximize the effect of rotation on the PMS, the initial rotation rate was taken to be the upper part of the distribution of rotation period observed in young open clusters \citep{GalletBouvier2015}. It corresponds to an initial rotation rate $\Omega_{\rm ini} = 7.3\, \mu\text{Hz}$ \citep{AmardEtal2016}.}

The radial profiles of the parameters relevant to our study, obtained from these models, are shown on Fig. \ref{fig:models} for ages ranging from 10 Myr to 4.6 Gyr, as a function of radius normalised by the one of the star at each specific age $t$: $R_*(t)$. The ZAMS occurs at 56 Myr in our model.  {On both panel, dashed lines correspond to profiles on the PMS, and solid lines correspond to profiles on the MS.}
The left panel of Fig. \ref{fig:models} shows the Richardson number Ri  {with the definition above}. It can be seen that $\text{Ri} \gg 1$ during the whole evolution. Thus, because this parameter appears as $\text{Ri}^{-1/2}$ in the expression of the penetration length given by Eq. (\ref{eq:L}), we do not expect this term to play an important role. 
On the right panel of Fig. \ref{fig:models},  {we show } 
the ratio $N/2\Omega$. As can be seen on the latter, during the PMS
${\max(N/2\Omega)} \sim \text{a few}$, while this ratio builds up to reach $\sim 10$ at 250 Myr, and $\sim 100$ at 4.6 Gyr.  {This is mainly due to the fact that $N$ is increasing along the PMS, as the radiative core grows. In addition,} the global rotation rate of the star decreases along the MS due to wind braking. Thus, because this parameter appears as $2\Omega/N$ in Eq. (\ref{eq:L}), we expect it could play a role during the PMS, while during the MS we would have $2\Omega/N \ll 1$.

\subsection{Results}
\begin{figure}
\vspace{10pt}
\centering
\begin{tikzpicture}[scale=1.0/0.98]
     \node[anchor=south west,inner sep=0] at (0,0) {\includegraphics[width=\linewidth]{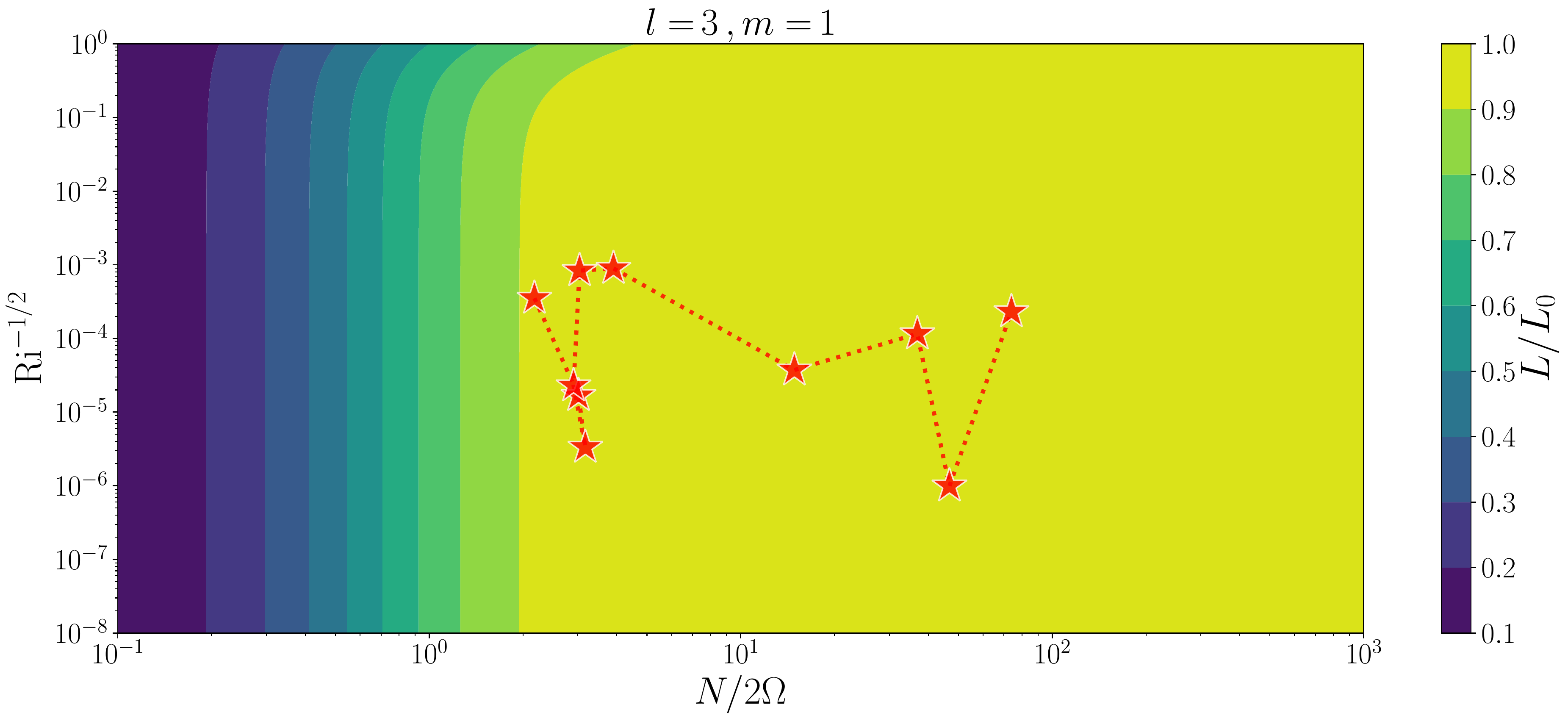}};
     \draw (7.44,0.52) -- ++ (0.96,0);
     \draw (0.12,3.54) -- ++ (0,0.4);
     \draw[>-,color=red,  ultra thick] (1.1,0.0) -- node[below]{PMS $\left(\overline{N/2\Omega} \sim 1\right)$} (6.35,0);
     \draw[->,color=blue, ultra thick] (6.41,0.0) -- node[below]{MS  $\left(\overline{N/2\Omega} \gg  1\right)$} ++ (8.3,0);
     \draw[dotted,color=white, ultra thick] (6.38,0.95) -- ++ (0,6.3);
     \node at (5.15,4.48)  {\footnotesize \texttt{10Myr}};
     \node at (5.52,3.55)  {\footnotesize \texttt{20Myr}};
     \node at (5.75,2.7)   {\footnotesize \texttt{30Myr}};
     \node at (6.75,3.38)  {\footnotesize \texttt{40Myr}};
     \node at (5.9,5.05)   {\footnotesize \texttt{50Myr}};
     \node at (6.9,5.07)   {\footnotesize \texttt{92Myr}};
     \node at (8.5,3.4)    {\footnotesize \texttt{250Myr}};
     \node at (9.7,4.4)    {\footnotesize \texttt{550Myr}};
     \node at (10.15,2.18) {\footnotesize \texttt{1Gyr}};
     \node at (11.2,4.59)  {\footnotesize \texttt{4,6Gyr}};
\end{tikzpicture}\\[-10pt]
\caption{Penetration length of gravito-inertial waves superimposed with the  {evolution}
track of a  {1 M$_\odot$ star}
, as a function of $\overline{N/2\Omega}$ and $\overline{\text{Ri}}^{-1/2}$, for $l=3$ and $m=1$. The colors indicate the penetration length of gravito-inertial waves $L$, given by Eq. (\ref{eq:L}), normalised by that of pure gravity waves, $L_0$. The red stars indicate the location of the calculated sun-like star in this  {parameters' space}
, calculated by averaging the profiles shown on Fig. \ref{fig:models} over the radiative zone.}
\label{fig:Sun_radius}
\end{figure}
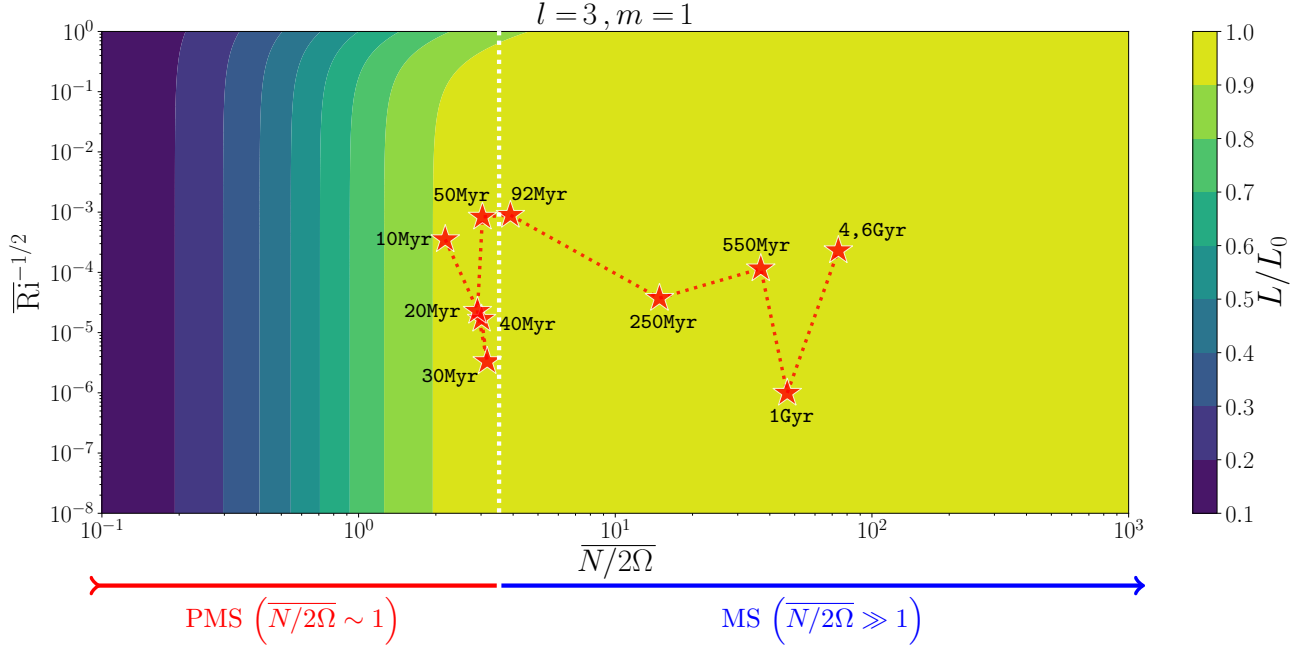
We now show how our new prescription for the penetration length, given by Eq. (\ref{eq:L}), depends on the control parameters $\text{Ri}^{-1/2}$ and $N/2\Omega$. Because these parameters are complicated functions of radius, as shown in the previous section, we plot the ratio $L/L_0$ as a function of their average values in the star, defined as 
$$
\bar{x} = \int_{\text{RZ}} x(r)\, \frac{\text{d}r}{R_{\text{RZ}}},
$$
where ${R_{\text{RZ}}}$ is the size of the radiative core (that can vary with age). This is shown on Fig. \ref{fig:Sun_radius} thanks to the color contours, which indicate the magnitude of the ratio $L/L_0$ for $l=3$, $m=1$ and $\text{Fr} = 0.01$. The penetration length $L$ (with rotation) is always lower than the one for pure gravity waves in our parameters range. In addition, one can see that $L/L_0$ mainly varies with the ratio $N/2\Omega$, the dependence as a function of $\text{Ri}^{1/2}$ being very shallow except when this parameter is close to unity, which is never the case in our solar-like model.

On Fig. \ref{fig:Sun_radius}, we have  {superimposed }
the evolutionary track of the model introduced above, in this parameter plane. This is shown by the red stars, with corresponding ages printed next to them. The vertical dotted line indicates the value of the ratio $\overline{N/2\Omega}$ at the ZAMS (56 Myr in our model).
The penetration length of gravito-inertial waves corresponding to radial averages of Ri and $N/2\Omega$ is close to is close to the one of pure gravity waves  {with at most 10\% difference during the early PMS.}

{However, we still expect that the penetration length could be significantly altered by rotation at specific locations, typically where the buoyancy frequency $N \rightarrow 0$.}
\begin{figure}
     \centering
     \includegraphics[width=0.495\linewidth]{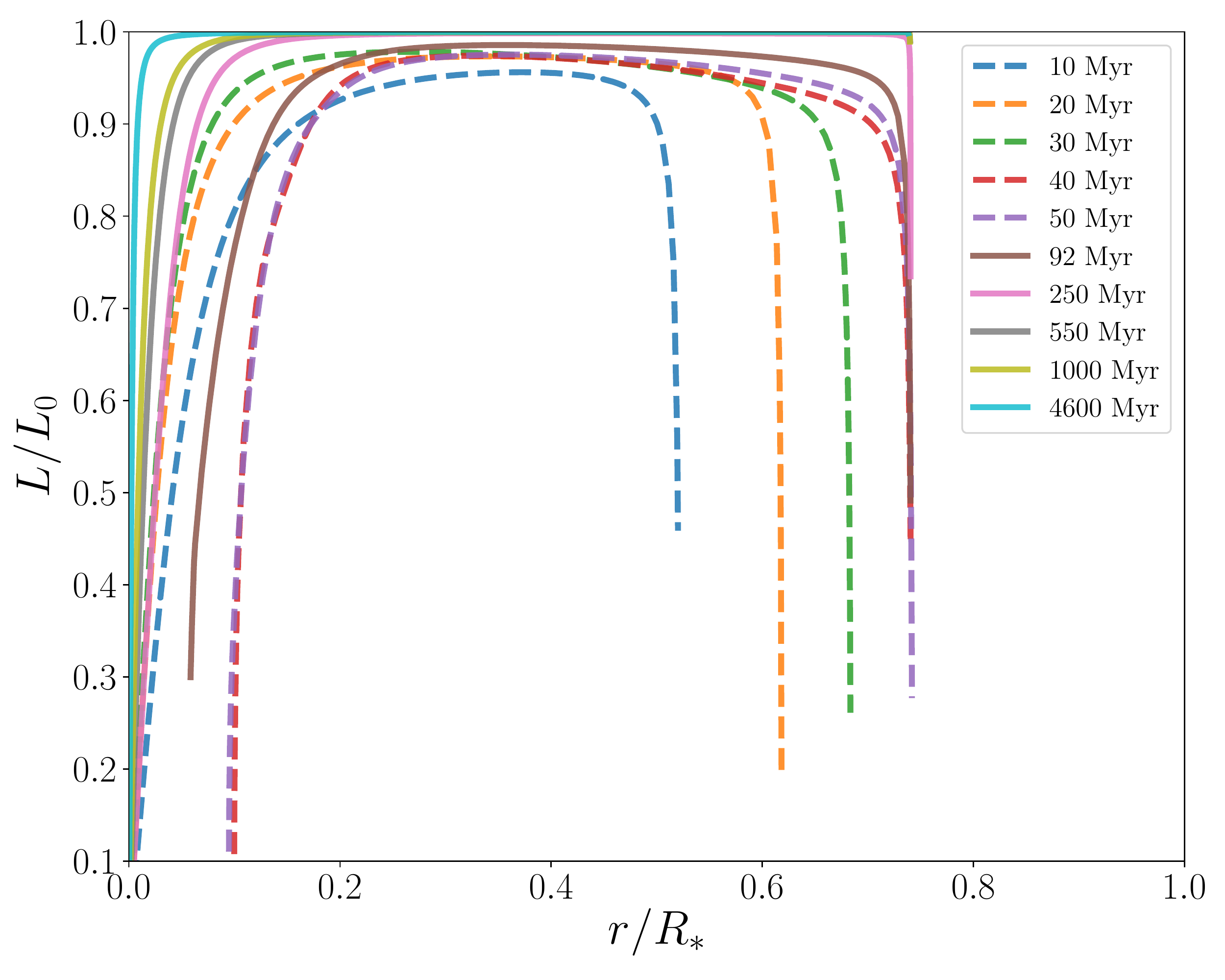}\\[-10pt]
     \caption{Normalised penetration length calculated from Eq. (\ref{eq:L}), as a function of the radius normalised by the stellar radius $R_*$ at specific ages, scanning from 10 Myr to 4.6 Gyr, obtained with the stellar evolution code STAREVOL. The ZAMS occurs at 56 Myr in the simulation. Dashed lines correspond to profiles on the PMS, and solid lines correspond to profiles on the MS}
       \label{fig:Sun_L}
\end{figure}
{To explore this, we have plotted on Fig. \ref{fig:Sun_L} the radial dependence of the ratio $L/L_0$, calculated from Eq. (\ref{eq:L}), for our different models presented on Fig. \ref{fig:models}. It can be seen that the penetration length of gravito-inertial waves is significantly descreased by rotation, near both boundaries of the radiative zone. This is where the buoyancy frequency smoothly matches the small negative value of the nearby turbulent convection zone. Thus, gravito-inertial waves excited by convective penetration or overshoot, that are produced in the matching region between convective and radiative zones, are expected from this analysis to be strongly impacted in their excitation region. In addition, we see on Fig. \ref{fig:Sun_L} that the penetration length is significantly decreased near the core of sun-like stars, where the buoyancy frequency also goes to zero.}

{As a result, we expect that the properties of propagation, dissipation, and associated angular momentum transport by gravito-inertial waves to be strongly impacted in those area.}

\section{Conclusions}
\label{sec:sec3}

{We have derived an analytical expression of the wave penetration length in an equatorial plane, including global and differential rotation.} We found that the penetration length of gravito-inertial is not significantly modified by rotation on average in the case {of} a solar-like star, even during the PMS where the ratio $\overline{N/2\Omega}$ is the lowest. We expect the same to be true for any low-mass star, as the ratio $N/2\Omega$, which we have shown to be the main control parameter {of the problem}, is expected to be large in this case. {However, we understood from our analysis that the wave penetration length can be strongly affected by rotation near their excitation region, and near the center of sun-like stars.}

We conclude that the prescriptions for angular momentum  {deposit because of the thermal damping of} pure gravity waves, implemented in state-of-the-art 1D stellar evolution codes, should be robust to the presence of (differential) rotation in the case of low-mass stars, {except in narrow regions where the buoyancy frequency $N \rightarrow 0$}.  {However, the picture will be more complex since the stochastic excitation of the waves can be strongly affected by rotation \citep{MathisEtal2014,Rogers2015}, with potential impact upon the resulting rotation profile.}

Moreover, as we consider higher mass stars, we think that rotation could play a more prominent role in this analysis. In the case of $\gamma$-doradus and Be stars for example, which undergo very rapid rotation, we expect that the penetration length of internal waves could be significantly decreased{, even in the bulk of radiative zones}. This should have important consequences for the understanding of angular momentum transport by internal waves in intermediate-mass and massive stars in general, and for the characterization of g-modes in O and B stars observed by Kepler. These will be the focus of attention of follow-up studies.

\begin{acknowledgements}
Acknowledgements: QA and SM acknowledge funding by the European Research Council through ERC SPIRE grant 647383, and the PLATO CNES grant at CEA-Saclay. LA also thanks the ERC through grant 682393 (AWESoMeStars).
\end{acknowledgements}

\bibliographystyle{aa}  
\bibliography{stars} 

\end{document}